\documentclass[%
aps, twocolumn, suprescriptaddress,
superscriptaddress,
showpacs, preprintnumbers,
 amsmath,amssymb,prl
]{revtex4-2}

\usepackage{graphicx}
\usepackage{dcolumn}
\usepackage{bm}
\usepackage{amsmath}
\usepackage{hyperref}
\usepackage{epstopdf}

\usepackage[mathlines]{lineno}

\usepackage{color}
\usepackage{ulem}
\usepackage{soul}

\usepackage{array}
\usepackage{makecell} 

\newcommand{\ket}[1]{$\left\vert #1\right \rangle$}
\newcommand{\bra}[1]{$\left\langle #1\right \vert$}
\newcommand{\un}[2]{$#1\,\mathrm{#2}$}
\newcommand{\psirel}{\psi^\mathrm{RM}}
\newcommand{\phicm}{\phi^\mathrm{CM}}
\newcommand{\psirelt}{\psi^\mathrm{RM}_\mathrm{t}}
\newcommand{\phicmiso}{\phi^\mathrm{CM}_\mathrm{iso}}

\usepackage{xcolor}

\begin{document}

\title{Observation of confinement-induced resonances in a 3D lattice}

\author{Deborah Capecchi}
 \affiliation{Institut f\"ur Experimentalphysik und Zentrum f\"ur Quantenphysik, Universit\"at Innsbruck, 6020 Innsbruck, Austria.}

\author{Camilo Cantillano}%
\affiliation{Institut f\"ur Experimentalphysik und Zentrum f\"ur Quantenphysik, Universit\"at Innsbruck, 6020 Innsbruck, Austria.}

\author{Manfred J. Mark}
\affiliation{Institut f\"ur Experimentalphysik und Zentrum f\"ur Quantenphysik, Universit\"at Innsbruck, 6020 Innsbruck, Austria.}
\affiliation{Institut f\"ur Quantenoptik und Quanteninformation, \"Osterreichische Akademie der Wissenschaften, 6020 Innsbruck, Austria.}

\author{Florian Meinert}\affiliation{5.Physikalisches Institut and Center for Integrated Quantum Science and Technology, Universit\"{a}t Stuttgart, 70569 Stuttgart, Germany.}

\author{Andreas Schindewolf}
\affiliation{Max-Planck-Institut f\"ur Quantenoptik, 85748 Garching, Germany.}
\affiliation{Munich Center for Quantum Science and Technology, 80799 M\"unchen, Germany.}

\author{Manuele Landini}
\affiliation{Institut f\"ur Experimentalphysik und Zentrum f\"ur Quantenphysik, Universit\"at Innsbruck, 6020 Innsbruck, Austria.}

\author{Alejandro Saenz}
\affiliation{Institut f\"ur Physik, Humboldt-Universit\"at zu Berlin, 12489 Berlin, Germany.}

\author{Fabio Revuelta}
\affiliation{Grupo de Sistemas Complejos, Escuela T\'ecnica Superior de Ingenier\'ia Alimentaria y de Biosistemas, Universidad {Polit\'ecnica} de Madrid, 28040 Madrid, Spain.}

\author{Hanns-Christoph N\"agerl}
\affiliation{Institut f\"ur Experimentalphysik und Zentrum f\"ur Quantenphysik, Universit\"at Innsbruck, 6020 Innsbruck, Austria.}


\date{\today}

\begin{abstract}	
We report on the observation of confinement-induced resonances for strong three-dimensional (3D) confinement in a lattice potential. Starting from a Mott-insulator state with predominantly single-site occupancy, we detect loss and heating features at specific values for the confinement length and the 3D scattering length. Two independent models, based on the coupling between the center-of-mass and the relative motion of the particles as mediated by the lattice, predict the resonance positions to a good approximation, suggesting a universal behavior. Our results extend confinement-induced resonances to any dimensionality and open up an alternative method for interaction tuning and controlled molecule formation under strong 3D confinement.
\end{abstract}

\maketitle

Cold atoms are a well-established platform for quantum simulation \cite{Gross2017qsw} and quantum computation \cite{Garcia-Ripoll2005}. They
allow for the realization of strongly-correlated quantum many-body phases \cite{Bloch2008mbp} and promise the investigation of dynamical processes in correlated quantum matter \cite{Mitra2018qqd} with exquisite parameter control. The cold-atom setting comes with two advantageous and enabling features: External potentials that can be flexibly created by optical means, e.g.~optical lattices~\cite{Morsch2006, Windpassinger2013eno}, and interatomic interactions that can be tuned almost at will, via magnetic Feshbach resonances (FRs)~\cite{Chin2010}. These features have been instrumental to a multitude of spectacular results, e.g.~the observation of the superfluid-to-Mott-insulator quantum phase transition~\cite{Greiner2002,Dutta2015nsh} and the Bose-Einstein-condensation-to-Bardeen-Cooper-Schrieffer (BEC-to-BCS) crossover~\cite{Strinati2018tbb}. Besides interaction tuning, FRs serve as an entrance door to the world of ultracold molecules~\cite{Regal2003cum,Herbig2003poa}, with great promises for ultracold chemistry~\cite{Balakrishnan2016pum}, precision measurement~\cite{Zelevinsky2008pto}, and dipolar many-body physics~\cite{Baranov2012cmt}. For a FR to occur, the free scattering state of two atoms needs to be brought to degeneracy with a molecular bound state. In most applications, a variable external magnetic field, as a result of the Zeeman effect, tunes the energy difference, inducing the resonance.

Tight external confinement as provided by, e.\,g., a lattice can also lead to a dramatic modification of the atoms' scattering properties \cite{Olshanii1998,Petrov2000,Bergeman2003,Fedichev2004,Peano2005,Grishkevich2009,Buechler2010,Sala2012}. Confinement-induced resonances (CIRs) occur if the typical length scale of the confining potential, e.\,g., the harmonic-oscillator length $a_\mathrm{h}$, and the 3D s-wave scattering length $a_\mathrm{s}$ assume similar values. CIRs come in two flavors: Elastic CIRs (ECIRs) emerge if the effective 1D or 2D interaction strength diverges at a particular ratio of these two length scales \cite{Olshanii1998,Petrov2000,Bergeman2003}. They are especially suited for tuning the interaction strength. Inelastic CIRs (ICIRs) \cite{Sala2012,Sala2016,Schulz2015} emerge due to the coupling between the relative motion (RM) and the center-of-mass motion (CM) \cite{Bolda2005,Peano2005,Schneider2009,Kestner2010}. At ICIRs, which form an infinite series, atoms can be transferred to the RM bound state after the scattering process, making such resonances more prone to losses, but also enabling molecule formation \cite{Sala2013}, similar to magnetic FRs. For ECIRs and ICIRs to occur, scattering properties can be tuned by variation of either $a_\mathrm{s}$ or $a_\mathrm{h}$.


\begin{figure}
	\includegraphics[width=\linewidth]{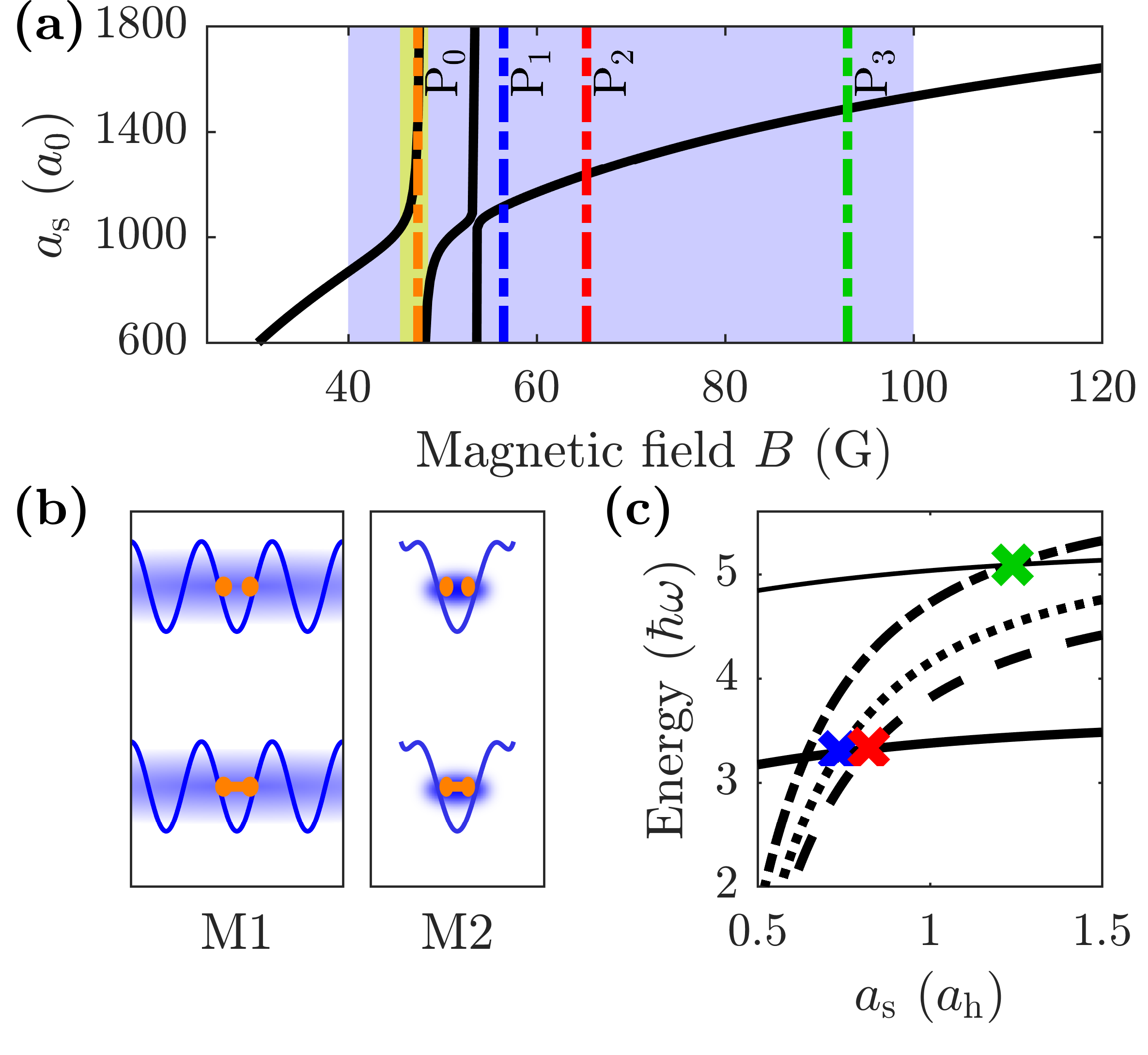}
	\caption[Fig:1]{(a) Scattering length $a_\mathrm{s}$ for Cs $(F,m_F)\!=\!(3,3)$ as a function of $B$ (solid black line) \cite{Berninger2013}, with two narrow FRs at \un{47.8}{G} and \un{53.8}{G}. The purple (yellow) shaded area corresponds to the interval used in experiment E1 (E2). The dashed lines labeled $P_0$, $P_1$, $P_2$, and $P_3$ mark the positions of the resonant features observed in E1. (b) Schematic representation of the states involved in the ICIRs: trap (upper) and bound (lower) state according to each model of the atom pair \cite{Sala2012}. For model M1 (M2) the trapping potential is given by the lattice (sextic) potential. Initially, both atoms are in the same lattice site in a trap state (separated orange circles), while the CM (blue area) can be spread (M1) or localized (M2). ICIRs occur if the anharmonicity of the trapping potential couples the trap state to the least-bound state (connected orange circles) with some center-of-mass excitations. (c) Energy diagram for varying $a_\mathrm{s}$ in units of $a_\mathrm{h}$ for isotropic trapping. The thick and thin solid lines correspond to $\psirelt \phicmiso(0,0,0)$ and $\psirelt \phicmiso(2,0,0)$, respectively, whereas the dashed, dotted, and dashed-dotted curves represent $\psi_b^\mathrm{RM} \phicmiso(4,0,0)$, $\psi_b^\mathrm{RM}\phicmiso(2,2,0)$, and $\psi_b^\mathrm{RM}\phicmiso(6,0,0)$, respectively. Intersections causing the ICIRs of the present work are indicated by crosses.}
    \label{Fig:1}
\end{figure}


Experimentally, so far, an ECIR in 1D has been used to put samples of ultracold bosons deeply into the Tonks-Girardeau regime and to access the super-Tonks-Girardeau (sTG) state \cite{Haller2009}. For spin-mixtures of fermions, the sTG state has been realized in the limit of two particles by means of an ECIR \cite{Zuern2012}. Tuning on the repulsive side of an ECIR has been used to map out three-body correlations in the strongly interacting 1D regime \cite{Haller2011tbc}, while in 2D the existence of an ECIR for attractive interactions has been confirmed \cite{Froelich2011}. Recently, the interplay of contact and dipolar interactions in conjunction with ECIRs has allowed the creation of scar states \cite{Kao2021tpo}. The observation of loss features in the 1D-to-2D crossover regime \cite{Haller2010} was originally interpreted as stemming from the ECIR and later recognized to originate from ICIRs \cite{Peng2011, Sala2012}. ICIRs have been observed in mixed dimensionality \cite{Lamporesi2010} and molecules have been coherently produced on ICIRs in 1D \cite{Sala2013}.


While the work discussed so far was limited to 1D or 2D systems, in this Letter we experimentally detect ICIRs in 0D as a result of strong 3D confinement and compare our data to theory. Interestingly, we detect the resonances starting from a Mott-insulator state with predominantly single-site occupancy, a situation for which one would expect at first glance a strong suppression of atom-atom scattering. We employ Cs atoms in the hyperfine $(F,m_F)\!=\!(3,3)$ ground state, which features wide tunability for $a_s$ due to a combination of broad and narrow magnetic FRs \cite{Berninger2013}, as shown in Fig.~\ref{Fig:1}(a). We take data with two different experimental setups, one (E1) \cite{Reichsoellner2017} for tuning $a_s$ on a broad background, and the other (E2) \cite{Mark2017} for tuning near a comparatively narrow FR.

Our theoretical methods build on previous models employed for the prediction of ICIR positions \cite{Sala2012,Sala2016}. We study our data by two complementary approaches, named in the following M1 and M2 \cite{SuppMat}. Both descriptions provide an approximate solution of the two-body problem in the presence of the 3D-lattice potential.

The potential energy of two atoms is given by $V(\textbf{r}_1,\textbf{r}_2)=V_{\mathrm {RM}}(\textbf{r})+V_{\mathrm{CM}}(\textbf{R})+W(\textbf{r},\textbf{R})$, where $\textbf{r}_1$ and $\textbf{r}_2$ are the positions of atoms 1 and 2, while $ \textbf{r}$ and $\textbf{R}$ are the relative and center-of-mass coordinates, respectively. For M1, $V_{\mathrm{RM}}$ is Taylor expanded to order $r^6$ (sextic potential), and the eigenenergies of the RM are found perturbatively employing the solution for two particles in a harmonic trap interacting via a pseudo-potential \cite{Busch1998}. The CM is solved exactly, considering the form of the lattice potential. In particular, M1 accounts for the full lattice bandwidth, without imposing any restriction on the quasimomentum, resulting in an allowed range for each crossing between the levels associated with an ICIR rather than in a fixed condition~\cite{SuppMat}. Finally, the coupling term $W$, which originates from the anharmonicity, is treated perturbatively, leading to avoided crossings. For M2, the trap-potential terms are expanded in the form of sextic potentials and the solution is found numerically, with the interaction between the atoms being described by an \textit{ab-initio} potential \cite{SuppMat}. The sextic Taylor expansion employed in both models for the RM implies that the two particles occupy the same well. M1 (as opposed to M2) can describe the motion of the atoms moving as pairs through the lattice, as illustrated in Fig.~\ref{Fig:1}(b) \footnote{The numerical simulations for M1 have been carried out using Cesium atoms, while those for M2 have been conducted with Lithium, whose validity has been previously assessed by comparison with a Cesium experiment \cite{Sala2012}.}.

%

In both models the RM eigenfunction can be classified by two families of states of different character. On the one hand, there are molecular-bound states,
the most loosely (least) one being denoted as $\psirel_{\rm b}$. On the other hand, there are higher-lying unbound states
strongly influenced by the external trapping potential. These states energetically lie in the dissociation continuum in the absence of the external potential, and become discrete trap states in a few-well potential or bands in a periodic lattice.
The lowest-lying one is denoted as~$\psirelt$. The function $\phicm(n_x,n_y,n_z)$ describes the solutions for the CM of the atom pair in the three directions with $n_i=0,1,2,...$ \cite{SuppMat}. In the isotropic lattice the excited bound states are typically
three-fold degenerate because of the factorization of the potential along the three directions. In the following, we will denote the degenerate states with the compact notation $\phicmiso(n_j,n_k,n_l)$
implying
any permutation of $\{n_x,n_y,n_z\}$. A schematic drawing of the relevant states for the reported measurements can be found in Fig.~\ref{Fig:1}(c), where we plot the energy as a function of $a_\mathrm{s}$ in units of $a_\mathrm{h}=\sqrt{\hbar/(m\omega)}$, with $\omega$ the
harmonic trapping frequency of the lattice in the isotropic case. The results are similar for both models. Additional crossings that violate parity conservation and cannot give rise to RM-CM coupling are not included in Fig.~\ref{Fig:1}(c).




The experimental procedures are similar for both set\-ups, and we give details here only for E1. We start from a BEC of $\sim$3$\times10^{4}$ $^{133}$Cs atoms \cite{Weber2003bec,Kraemer2004} with a condensate fraction of 70 to 80\%, levitated against gravity by a
magnetic field gradient $\left\vert\nabla{B}\right\vert \approx 31$ G/cm and held in a crossed optical dipole trap (XODT) at $\lambda=1064.5$ nm by trapping beams along the horizontal $x$ and $y$ directions. The trap depth of the XODT is $V_{\textrm{trap}}\!=\! k_\text{B} \times 0.29$ $\mathrm{\mu K}$ with trapping frequencies $\nu_x=10.6(1.2)$ Hz, $\nu_y=16.0(1.7)$ Hz, and $\nu_z=20.8(1.5)$ Hz, where $z$ denotes the vertical direction along gravity.

\begin{figure}
\includegraphics[width=\linewidth]{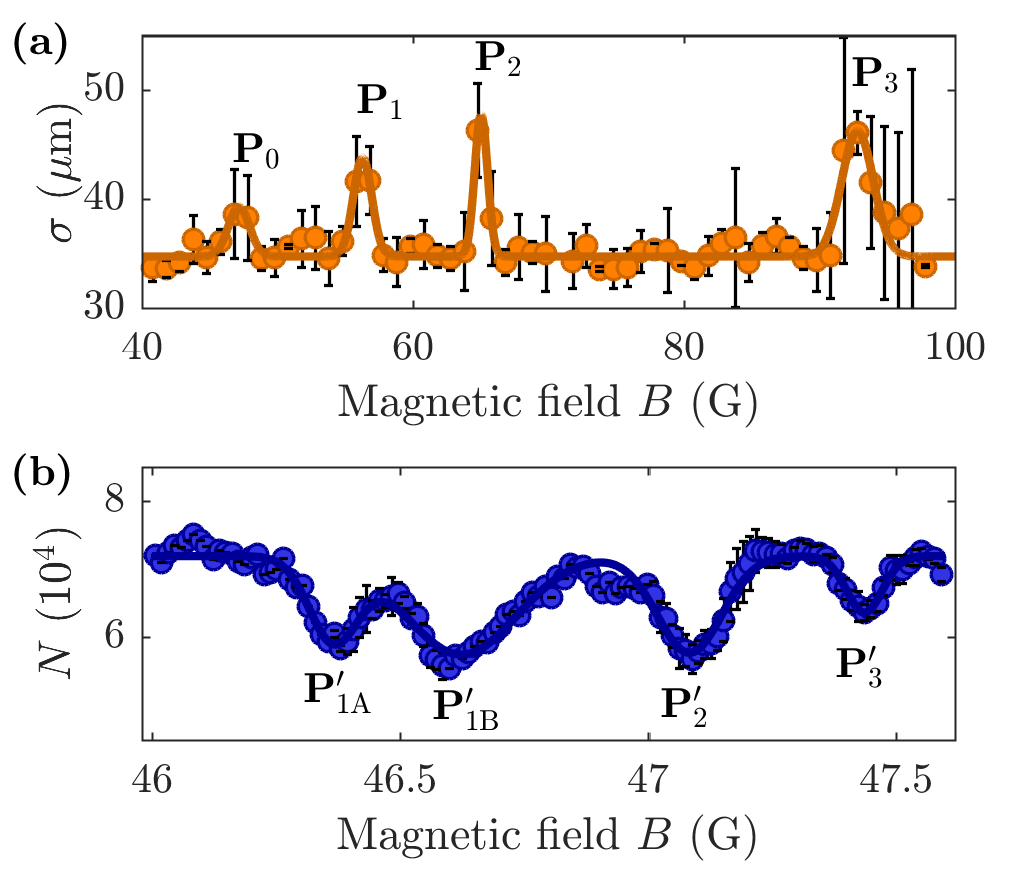}
\caption[Fig:2]{Observation of ICIRs in a 3D lattice. (a) Cloud radius $\sigma$ as a function of $B$ for E1 at $V_{x,y,z}=20.0(3)$ $E_\mathrm{R}$. The data are fit by multi-peak Gaussian (solid line) to guide the eye. (b) Atom number $N$ as a function of $B$ for E2, showing four loss features $P'_j$, $j=$ 1A, 1B, 2, 3. The solid line is also a multi-peak Gaussian fit. The errors bars in both plots reflect the standard deviation from typically 3 experimental runs.} \label{Fig:2}
\end{figure}

We load the BEC into a 3D cubic lattice with a depth of up to $V_{x,y,z}\!=\!20.0(3)$ $E_\mathrm{R}$ along all three directions, where $E_\mathrm{R}$ is the Cs recoil energy, by ramping up the power in the three retro-reflected lattice beams with wavelength $\lambda$ within \un{500}{ms}. Together with a fine adjustment of the XODT to control the chemical potential this creates a Mott insulator with predominant single-site occupancy \cite{Reichsoellner2017,SuppMat}. The offset magnetic field $B$ is subsequently increased from $B_\textrm{i}\!=\!21.0\,$G to a value $B_\textrm{0}$ between \un{40}{G} and \un{100}{G}, while adjusting $\left\vert\nabla {B}\right\vert$ to keep the atoms magnetically levitated. In this interval, $a_s$ varies in the range from $\sim$1.0$\times 10^3\,a_0$ to $\sim$1.5$\times10^3\,a_0$, where $a_0$ is the Bohr radius, skipping the narrow FRs. We hold the atoms in the lattice for a time $\tau_{\textrm{h}}=15$ to $20$ ms, after which we reverse the procedure above. We image the atoms after \un{52}{ms} time of flight and we record the atom number $N$ and the rms radius $\sigma$ of the sample.

\begin{figure}
    \includegraphics[width=\linewidth]{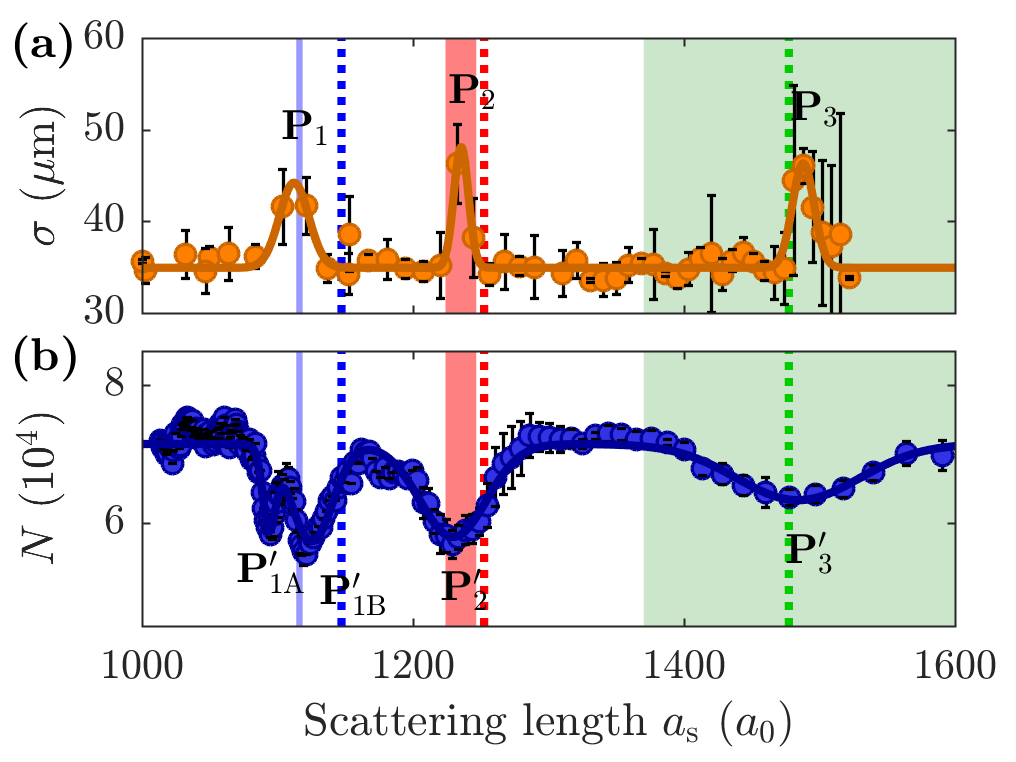}
    \caption[Fig:3]{Comparison of the data from the two experiments. (a) Cloud radius $\sigma$ from E1 and (b) atom number $N$ from E2 as a function of $a_\mathrm{s}$. The resonance positions $a_r$ are found at $1112(22)$ $a_0$, $1236(11)$ $a_0$, and $1488(16)$ $a_0$ for $P_1$, $P_2$, and $P_3$, respectively, and at $1094(10)$ $a_0$, $1124(26)$ $a_0$, $1233(40)$ $a_0$, and $1484(92)$ $a_0$ for $P'_\mathrm{1A}$, $P'_\mathrm{1B}$, $P'_\mathrm{2}$, and $P'_\mathrm{3}$, respectively. For each peak, the resonance position is obtained as the center of a Gaussian fit, and the Gaussian sigma is indicated in the parenthesis. The colored areas indicate the intervals obtained from M1, and the dotted lines represent the positions of the crossings as given by M2: $\psirelt \phicm(0,0,0)$ with $\psi_b^\mathrm{RM}\phicmiso(2,2,0)$ (blue) and with $\psi_b^\mathrm{RM} \phicmiso(4,0,0)$ (red), and $\psirelt \phicmiso(2,0,0)$ with $\psi_b^\mathrm{RM} \phicmiso(6,0,0)$ (green). The width of the blue area has been increased by a factor of 20 to improve visibility.} \label{Fig:3}
\end{figure}

With E1 we observe 4 characteristic heating features. The data are shown in Fig.~\ref{Fig:2}(a). We attribute the feature $P_0$ to the FR at 47.8 G. The resolution is not sufficient to allow for the observation of substructures near this FR. The much narrower FR at 53.8 G does not show up in the data. The features $P_1$, $P_2$, and $P_3$ at $ B\!=\!56.5(2)$ G, $65.3(3)$ G, and $93.0(2)$ G, respectively, do not correspond to known FRs. In fact, they are ICIRs, as is confirmed below. In E2, we tune $a_\mathrm{s}$ on the repulsive side of the FR at 47.8 G. This time we observe 4 atom-loss features $P'_\mathrm{1A}$, $P'_\mathrm{1B}$, $ P'_\mathrm{2}$, and $P'_3$ as $B$ is scanned from $46$ G to $47.6$ G as shown in Fig.~\ref{Fig:2}(b). We convert $B$ into $a_\mathrm{s}$ according to Fig.~\ref{Fig:1}(a) and plot both data sets together in Fig.~\ref{Fig:3}. The positions of the heating features $P_1$, $P_2$, and $P_3$ agree reasonably well with the positions of the loss features $P'_\mathrm{1A}$ and $P'_\mathrm{1B}$, $ P'_\mathrm{2}$, and $P'_3$, respectively. As discussed below, in E2 the feature $P_1$ from E1 appears to be split into two components $P'_\mathrm{1A}$ and $P'_\mathrm{1B}$. We attribute the splitting to some slight anisotropy of the lattice in E2~\cite{SuppMat}. Features $P'_\mathrm{2}$ and $P'_3$ are significantly broader than $P_2$ and $P_3$, respectively. We attribute the broadening to the magnetic field gradient, which causes a spread of $a_s$ across the sample \cite{SuppMat}. This becomes increasingly larger as one climbs up the narrow FR at 47.8 G.

We now compare the experimental data with the predictions of the two models, without any fitting parameters.
Even though the models are quite different, they both reflect the data reasonably well. The positions of $P_1$ and $P_2$ are in good agreement with the
avoided crossings predicted by M2 of the state $\psirelt \phicm(0,0,0)$ with the three-fold degenerate states
$\psirel_b\phicmiso(2,2,0)$ and $\psirel_b\phicmiso(4,0,0)$, respectively. The difference in energy between the two states (2,2,0) and (4,0,0) stems from anharmonic corrections and hence leads to two different ICIRs. In Fig.~\ref{Fig:3} we indicate the calculated positions of the resonances from M2 and the intervals from M1. For $P_1$ and $P_2$ we again find reasonable agreement.

 It is interesting to note that, on the basis of our modelling, $P_3$ cannot arise from a crossing in which the state $\psirelt \phicmiso(0,0,0)$ is involved. However, this resonance could be the result of excitations in the Mott-insulator state. The avoided crossings of $\psirelt \phicmiso(2,0,0)$ with $\psirel_b\phicmiso(6,0,0)$, and $\psirelt \phicmiso(1,0,0)$ with $\psirel_b\phicmiso(5,0,0)$ are in fact very close to the position of $P_3$. We hypothesize that such excitations can be brought about by mixing of the lowest band with excited bands due to the strong interparticle interaction. In our case, the on-site interaction energy is $\sim$78\% of the bandgap to the first-excited lattice band~\cite{SuppMat}, which may open the route for a non-negligible population of the excited states, i.\,e., bands, in a fully correlated many-body description.
%


\begin{figure}
\includegraphics[width=1\linewidth]{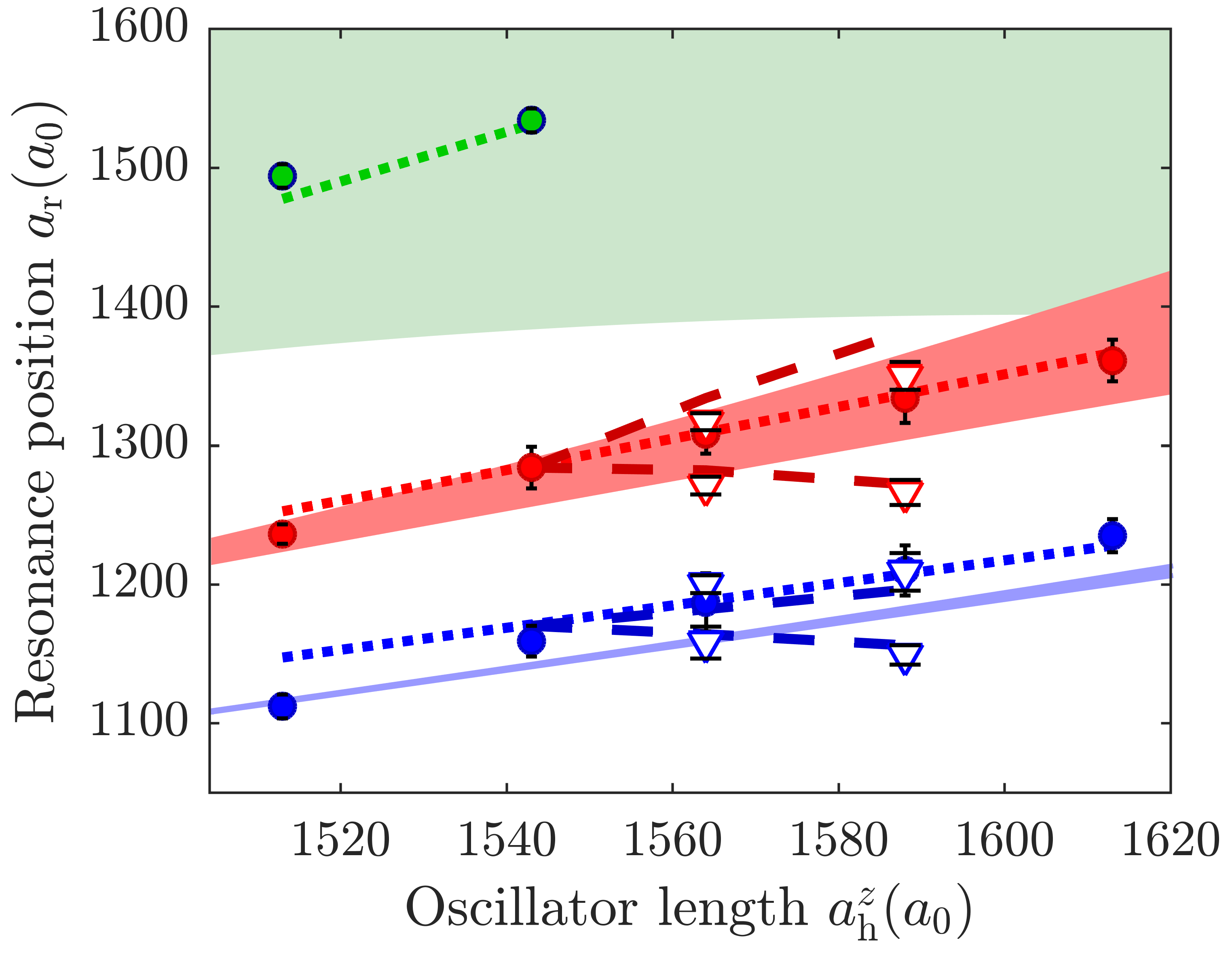}

\caption[Fig:4]{Tuning of the ICIRs. Resonance position $a_r$ for $P_1$ (blue), $P_2$ (red), and $P_3$ (green) as a function of the oscillator length $a^z_h$ along the $z$ direction for the isotropic (circles) and the anisotropic (triangles) case. For the isotropic case, $V_{x}\!=\!V_{y}\!=\!V_{z}$ was set to $20.0(3), 18.5(3), 17.5(3), 16.5(3)$, and $15.5(3)$ $E_\mathrm{R}$ for the data from left to right. For the anisotropic case, $V_x\!=\!V_y\!=\!18.5(3)$ $E_\mathrm{R}$ was chosen for $V_z\!=\!17.5(3)$ $E_\mathrm{R}$ (left triangles) and $16.5(3)$ $E_\mathrm{R}$ (right triangles). The error bars reflect the statistical uncertainties \cite{SuppMat}. The results from M1, applicable to the isotropic case, are shown as the blue, red, and green areas. The blue area has been widened by a factor of 3 to improve visibility. The predictions of M2 are plotted as dotted (dashed) lines for the isotropic (anisotropic) case. }
 \label{Fig:4}
\end{figure}

Next, we test how the stiffness of the lattice confinement and the introduction of some controlled anisotropy affect the resonances, i.\,e., how their positions are tuned and to what extent they split. Within E1 we first vary $V_{x,y,z}$ in the range between $20.0(3)$ and $15.5(3)$ $E_\mathrm{R}$ while maintaining $V_{x}\!=\!V_{y}\!=\!V_{z}$.  Model M1, which takes into account the full lattice in the treatment of the CM \cite{SuppMat}, allows us to provide a range~\footnote{The actual range is slightly larger due to the experimental uncertainties in, e.\,g., the laser intensities.}
within which the resonances are more likely to be found. This turns out to be very narrow for $P_1$ and increasingly larger for $P_2$ and $P_3$ due to the increasing bandwidth of the lattice bands. In Fig.~\ref{Fig:4} we plot the resulting resonance position $a_\mathrm{r}$, obtained in the same way as for the resonances shown in Fig.~\ref{Fig:3}, as a function of the harmonic oscillator length $a^z_\mathrm{h}=\sqrt{\hbar/(m \omega_z)}$ along the $z$ direction. As can clearly be seen for $P_1$ and $P_2$, the position shifts to higher values for $a_\mathrm{r}$ with increasing $a^z_\mathrm{h}$.
For $P_3$ only two data points are available, but also here a trend to higher values is present. For two specific values $V_z\!=\!17.5(3)$ $E_\mathrm{R}$ and $16.5(3)$ $E_\mathrm{R}$ we then set $V_x\!=\!V_y\!=\!18.5(3)$ $E_\mathrm{R}$. In such an anisotropic case, $P_1$ and $P_2$ split into two components,
similarly to what we had observed earlier with E2. The degeneracy is removed \cite{SuppMat} by increasing $a_\mathrm{h}^\mathrm{z}$ with respect to $a_\mathrm{h}^\mathrm{x,y}$. Here, the crossings of e.g.~$\psirelt \phicm(0,0,0)$ with $\psirel_{b}\phicm(2,0,2)$ and $\psirel_{b}\phicm(0,2,2)$ are found at the same position, but they are at a different position with respect to the crossing of $\psirelt \phicm(0,0,0)$ with $\psirel_{b}\phicm(2,2,0)$. For this reason, in the anisotropic lattice, each resonance splits into two separate resonances, with a separation that depends on the difference in the trapping potential along the three directions.

While in this work the positions and physical origin of the ICIRs, found experimentally to occur in strong 3D confinement, are well explained by the theoretical models, the details of how the resonance leads to loss and heating need further theoretical and experimental investigation.
In particular, the degree of suppression of the loss features in the Mott-insulator state and the possible role of inhomogeneities, allowing for superfluid regions in the system, is unclear. Studies of the loss dynamics and in particular of its density dependence would help elucidate the nature of the observed resonances, with connections to recent observations of tunnelling dynamics of doublons in the regime of strong interactions and under strong three-body losses \cite{Mark2020ibc}, something that is not considered in any of our theoretical models, which relies on previous observations of ICIRs in 1D geometry, where the loss was attributed to a two-step process via the creation of a mobile molecule \cite{Sala2013}. In  this context, detecting the ICIRs out of a two-atom Mott-insulator shell could greatly elucidate the relevant processes.


Our results contribute to the understanding and characterization of CIRs, with prospects for the association of dimers in optical lattices \cite{Meinert2016,Moses2015,Reichsoellner2017} and optical tweezers \cite{Wang2019,zhang2020} in the absence of, e.\,g., magnetic FRs, as a detection tool or as a step for creating ultracold molecules. In addition, these results add constraints for the stability of atoms in optical lattices and become very relevant in the presence of systems with high number of FRs like lanthanide atoms \cite{Patscheider2020, Kumlin2019}, or large
lattice depths as it is the case in quantum-gas microscopy \cite{Rispoli2019, Haller2015}.

We thank Erich Dobler for help with the experiment and discussions. The Innsbruck team gratefully acknowledges funding by the DFG-FWF Forschergruppe FOR2247 under the FWF project number I4343-N36, via a Wittgenstein prize grant under project number Z336-N36, and by the European Research Council (ERC) under project number 789017. Alejandro Saenz thanks that part of the research was supported by the National Science Foundation under Grant No. NSF PHY-1748958.
Fabio Revuelta acknowledges support by PGC2018-093854-BI00 and PID2021-122711NB-C21 funded by MCIN/AEI/10.13039/501100011033, and by
the Comunidad de Madrid under the Grant APOYO-JOVENES-4L2UB6-53-29443N (GeoCoSiM) financed within the
Plurianual Agreement with the Universidad Politécnica de Madrid in the line to improve the research of young doctors.

The data that support the findings of this study are made publicly 
available by the authors at \footnote{Dataset are available via Zenodo with doi:\url{10.5281/zenodo.7108194} }

\textit{Note added.}—We recently became aware of an independent
work by Lee \textit{et al.} \cite{Lee2022}.

\section{Supplemental Material}
This supplemental material provides further details for the results presented in the main text. For this purpose, first the adopted theoretical models are discussed. Second, a description of the two experimental settings is presented. Finally, we conclude with a brief discussion on the question of possible excitations to higher Bloch bands, which may explain the otherwise not assigned resonance P$_3$.
\subsection{Theoretical description}
The Hamiltonian for two atoms in the lattice potential is given by \cite{Sala2012}
 \begin{equation}\label{Hamiltonian}
\begin{aligned}
H(\textbf{r},\textbf{R})=&T_\mathrm{RM}(\textbf{r})+T_\mathrm{CM}(\textbf{R})+V_\mathrm{RM}(\textbf{r})\\
&+V_\mathrm{CM}(\textbf{R})+U_\mathrm{int}(\textbf{r})+W(\textbf{r},\textbf{R}),
\end{aligned}
\end{equation}
with $\textbf{r}=(\textbf{r}_1-\textbf{r}_2)/\sqrt{2}$ and $\textbf{\textbf{R}}=(\textbf{r}_1+\textbf{r}_2)/\sqrt{2}$ the coordinates for the relative motion (RM) and the center-of-mass motion (CM), respectively, and the interaction potential $U_\mathrm{int}$. The potentials for the RM and the CM take the form $V_\mathrm{RM}(\textbf{r})=2s E_\mathrm{R} \sum_{i}\sin^2\left(k r_i/\sqrt{2}\right)$ and $V_\mathrm{CM}(\textbf{R})= 2s E_\mathrm{R} \sum_{i}\sin^2(k R_i/\sqrt{2})$, while the coupling between the two motions is given by $W(\textbf{r},\textbf{R})=-4s E_\mathrm{R} \sum_{i}\sin^2\left(k r_i/\sqrt{2}\right)\sin^2(k R_i/\sqrt{2})$. Here, $s$ is the lattice depth in units of the recoil energy $E_\mathrm{R}$=$\frac{\hbar^2 k^2}{2m}$, $k$ is the wave number of the lattice light, and $i$ denotes $x$, $y$, or $z$.

To describe the system theoretically, we employ two complementary approaches, differing in the way the interatomic potential as well as the external potential are treated. A summary of the model differences can be found in Table$\ $\ref{Tab:2}.

\subsubsection{Interaction-potential treatment}
In the first approach (M1), we employ a pseudopotential approximation for the interatomic interaction potential, modeled by a regularized delta function $U_\mathrm{int}(r)=\frac{4\pi \hbar^2 a_s}{m}\delta(r)\frac{\partial}{\partial r}r$, where $a_s$ is the scattering length. The advantage of this approach is that most of the calculations can be carried out analytically. However, for the case of an anisotropic trapping potential, M1 cannot be applied due to a substantial complication of the computation.

The second approach (M2) deals with a more realistic interaction potential and is based on a full {\it ab-initio} calculation. The interaction potential with Cs atoms is, nevertheless, computationally challenging due to the large number of molecular bound states, which have, as a consequence, a more complex nodal pattern. In view of this, and assuming a universal regime for which the properties of the system can be entirely described by $a_\mathrm{s}$ and by the characteristic trap size $a_\mathrm{h}$, regime which has been assessed by comparison with the experiment reported in Ref.~\cite{Sala2012}, we perform the computation for two Li atoms, which have only 11 bound states and thus wavefunctions with much less nodes than the Cs$_2$ dimer.



\subsubsection{Trapping-potential treatment}
For the treatment of the external trapping potential, as an introductory step, we consider the case of an individual potential well in the harmonic approximation, for which relative motion (RM) and center of mass motion (CM) can be separated. For the RM, the eigenstates of the system feature one bound state $\psirel_{\textrm{b}}
$\footnote{for M1, this is the only bound state supported by the delta function. For M2, this is the bound state closest to threshold, other bound states are separated by a large energy gap} and an infinite ladder of trap-induced states $\psirel_n$, whose energies vary as a function of $a_\mathrm{s}$ \cite{Busch1998}. To these RM states, the CM adds a ladder of states $\phicm(n_x,n_y,n_z)$ separated by an integer number of excitations along the three trap directions $x, y$ and $z$. The term $W$, in our case caused by an anharmonicity of the trapping potential, can now provide the necessary coupling to create avoided crossings and hence to give rise to inelastic confinement induced resonances (ICIRs) \cite{Sala2012,Sala2016}.

The procedure followed in both models is based on the identification of the relevant crossings, as just explained for the case of the harmonic oscillator.

For both models, we initially neglect the coupling term $W$ and isolate the RM Hamiltonian and the CM Hamiltonian and calculate their eigenstates $\psirel$
and $\phicm(n_{x},n_{y},n_{z})$, respectively.

For M1, the potential $V_\mathrm{RM}$ is initially approximated by a quadratic function, with $\hbar\omega_\mathrm{RM}=2\sqrt{s}E_R$. For this problem, an exact solution is known \cite{Busch1998}. The energies are given implicitly by
\begin{equation}
 \frac{a_h}{a_s}=\frac{\sqrt{2}\Gamma\left(\frac{3}{4}-\frac{E_\mathrm{RM}}{2\hbar\omega_\mathrm{RM}}\right)}{\Gamma\left(\frac{1}{4}-\frac{E_\mathrm{RM}}{2\hbar\omega_\mathrm{RM}}\right)}
 \end{equation}
 with $a_h=\sqrt{\frac{\hbar}{m \omega_\mathrm{RM}}}=\frac{1}{k s^{1/4}}$. The corresponding wavefunctions are given by
 \begin{equation}
 \begin{aligned}
 \psirel \left(\textbf{r}\right)=&\frac{1}{2\pi^{3/2}}A e^{-\frac{r^2}{2 a_h^2}} \Gamma\left(\frac{3}{4}-\frac{E_\mathrm{RM}}{2\hbar\omega_\mathrm{RM}}\right)\times\\
 &\mathcal{U}\left(\frac{3}{4}-\frac{E_\mathrm{RM}}{2\hbar\omega_\mathrm{RM}},\frac{3}{2},\frac{r^2}{a_h^2}\right),
 \end{aligned}
 \end{equation}
where $A$ is a  normalization constant, $\mathcal{U}$ is the confluent hypergeometric function, and $\Gamma$ is the Gamma-function. We further calculate, by first order perturbation theory, corrections to the energies arising from anharmonicities of the form $r_i^4$ and $r_i^6$, as obtained by a Taylor expansion of $V_\mathrm{RM}$. In M1, the CM is solved exactly on the separable lattice $V_\mathrm{CM}$, with periodic boundary conditions \cite{Bolda2005}. In this case the wavefunctions $\phicm(n_{x},n_{y},n_{z})(\textbf{q})$ are Bloch functions with eigenenergies $ E^\mathrm{CM}_{n_{x},n_{y},n_{z}}(\textbf{q})$, where the indices $n_i$ take the meaning of band indices and $\textbf{q}$ is the quasi-momentum. The CM-RM coupling energy contribution is considered also as self-coupling on the CM and RM subspaces: \bra{\phicm}W\ket{\phicm} and similarly for $\psirel$.
In this way, we build a diagram similar to the one shown in Fig.~1(c) in the main text. The crossings are calculated for the two edges of the lattice band, $q=0$ and $q=k$. In the isotropic lattice, each resonance is three-fold degenerate due to the factorization of the CM in the three directions. 

For M2, the potential terms $V_\mathrm{RM}$, $V_\mathrm{CM}$ and $W$ are Taylor expanded up to powers of $r_i^6$ and the wavefunctions $\psirel$ and $\phicm(n_{x},n_{y},n_{z})$ are computed on a basis of
$B$ splines for the radial part and spherical harmonics for the angular part \cite{Grishkevich2009,Schneider2009}. In order to adequately reproduce the highly oscillatory inner part of the RM eigenfunctions (usually known as the molecular region), different knot sequences for the
$B$ splines are used. For the region close to the origin (up to a distance of 15 $a_0$) it is 50 $B$ splines following a linear distribution; in order to cover the smooth long-range part, 100 $B$ splines.
with a geometric distribution for larger distances are used
(up to 30,000$\sqrt{12}$\,$a_0$).
As the center-of-mass potential is a smooth function with no singularity at the origin, 100 $B$ splines uniformly distributed over $30,000\sqrt{3}a_0$ were found to be sufficient in the computation of the CM eigenfunctions. In order to calculate only those states that are totally symmetric, i.\,e.,
those states that belong
to the $A_g$ irreducible representation of the $D_{2h}$ symmetry group \cite{Grishkevich2009,Schneider2009}, the angular part of both the RM and the CM is described using the spherical harmonics with $\ell=0, 2, \ldots, 34$, and $m=0, 2, \ldots, 14$.

The eigenstates of the Hamiltonian in Eq.$\,$(\ref{Hamiltonian}) are then described in terms of configurations $\Phi_{i,j}$=$\psirel_{j}\phicm_{j}(n_{x},n_{y},n_{z})$, which are combinations of the RM and CM eigenfunctions, as detailed in Refs. \cite{Grishkevich2009,Schneider2009}. Then, the contribution from the coupling term can be calculated and the total Hamiltonian can be diagonalized. In M2, 200 eigenfunctions for the RM and 75 for the CM have been combined, rendering a basis set formed by 15000 elements. The simulation is performed in a lattice of \un{1064.5}{nm}, where the power of the beams is chosen such that the potential depth for Li in units of recoil is $s^{\mathrm{Li}}=s^{\mathrm {Cs}}$. The validity of the use of the interaction potential of Li atoms has been previously reported in the Ref.$\ $\cite{Sala2012} by explicit comparison with the Cs experiment described in the Ref. \cite{Haller2009}. The results are shown in both plots of Fig.~3 and Fig.~4 in the main text and are in fair agreement with the experimental data.
	\begin{table*}

        \setlength{\tabcolsep}{12pt}
 	\caption{Main characteristics of the models M1 and M2 that are used in the simulations.}
	\begin{tabular}{ l c  c c c}
	
		\hline
		\hline
	Characteristic & &M1 & &M2 \\
		\hline
		Optical trap && \makecell{Isotropic\\ (Anisotropic demanding)} & \qquad&\makecell{ Isotropic \& anisotropic} \\
		&&&&\\
	    RM&&&&\\
	    \quad a. Optical trap potential &&{Taylor expanded (order 6)}&& {Taylor expanded (order 6)} \\
		\quad b. Interaction potential & & $\delta$ pseudopotential &\qquad & {\it ab initio} (Li$_2$) \\
		\quad c. No. of bound (molecular) states && 1 && $11$ \\
        \quad d. Energies && Discrete levels /  Bands && Discrete levels / Bands  \\

		&&&&\\
		CM &&&&\\
		\quad a. Optical trap potential &&Full ($\sin^2$)&&Taylor expanded (order 6)  \\
		\quad b. Energies && Continuous bands && Discrete levels \\
		\hline
		\hline
	\end{tabular}

	\label{Tab:2}
	\end{table*}
 
Let us conclude by remarking that M2 allows also for Taylor expansions of higher order which reproduce multiwell setups. For example, a Taylor expansion of order 22 reproduces a triple-well system [54]. M2 is also able to describe a $\cos^2$ optical lattice, something that enables the study of double-well systems when Taylor expanded up to order 12. 
\begin{figure}
    \centering
    \includegraphics[width=\linewidth]{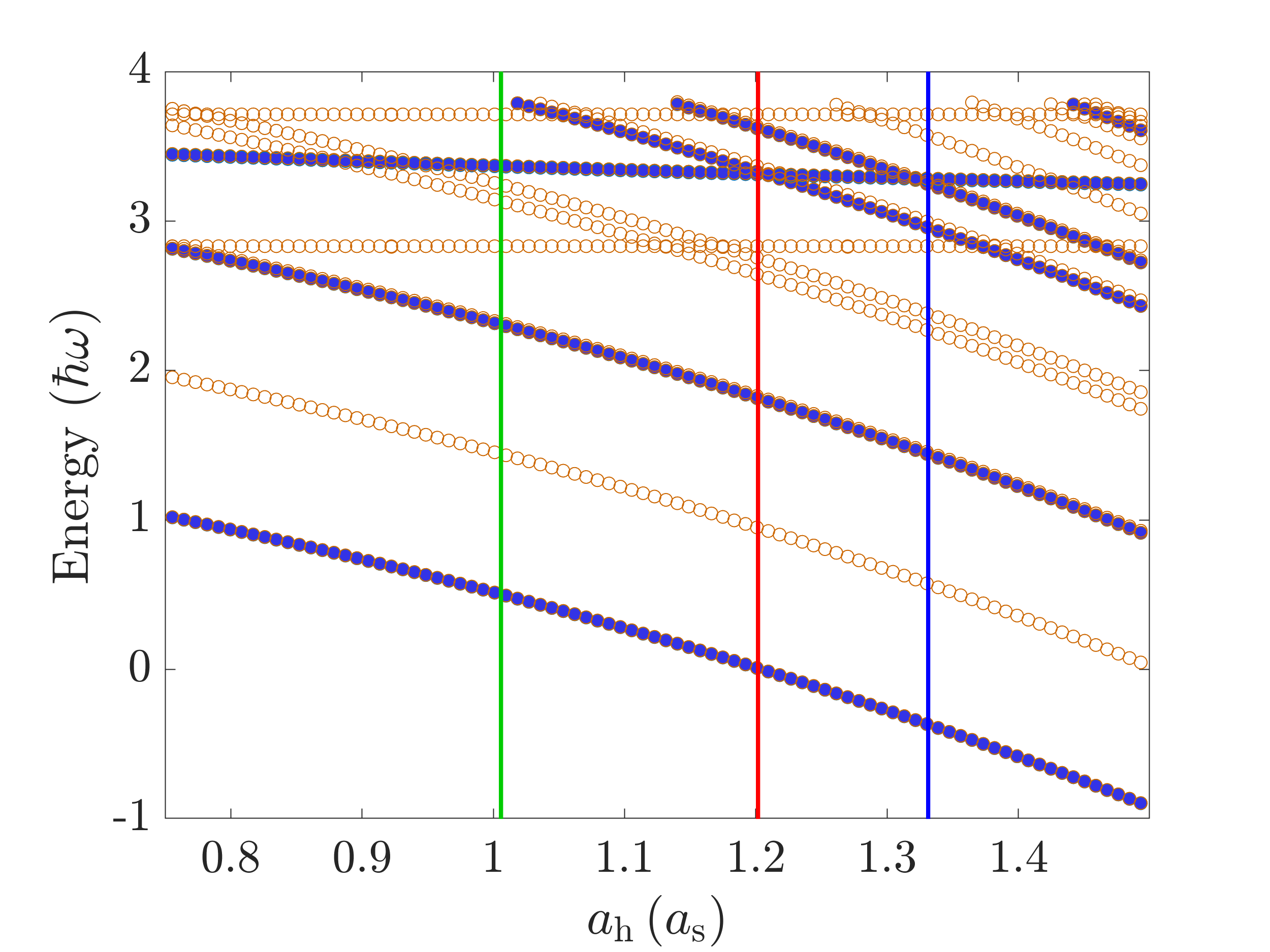}
    \caption{Correlation diagram for two Li atoms confined in an optical trap with $V_{x, y, z}=20 E_R$ formed by an isotropic single well (blue filled circles), and in a trap formed by a single well in two directions and a double well in the remaining one (orange empty circles). The vertical solid lines show the position of the experimentally observed confinement-induced resonance $P_1$, $P_2$, $P_3$, respectively in blue, red, and green.}
    \label{Fig:5}
\end{figure}

Fig.~\ref{Fig:5} shows the correlation diagram for two Li atoms interacting with an \textit{ab-initio} potential. The blue filled circles show the energy of the atoms when confined in a trap with a single well in all directions, while the orange empty circles show the energies for a trap with a single well in two of the spatial directions and a double well in the remaining one. As can be seen, the double-well system presents more energy levels than the single-well one. Some of them emerge due to the symmetry breaking in one direction (like those for the two molecular states on the right), while others are totally absent in the single well. That is the case for the first-trap state for the double well (bottom almost horizontal aligned points) whose energy is almost unaffected by the interaction strength between the two particles; this state corresponds to a situation where each of the atoms is found in a different well and then hardly interact due the short-range interparticle potential. Contrarily, the first-trap state for the single well strongly increases with the interaction strength as the atoms lie close to each other and then can repel much more strongly. Note, similarly, that this trap state is also present in the double well (but with a slightly larger energy). For vanishing interactions, like those used in the beginning of the experiment, the previous two trap states are degenerate, a fact that could explain why the observed resonances, whose positions are marked as vertical solid lines, involve the upper trap state instead of the bottom one, which is the one associated with the Mott-insulating state.
The comparison between the M2 single-well results has been assessed and the double-well calculations, renders differences of only a few atomic units in the ICIRs positions.

\subsection{Experimental procedures}

\subsubsection{Details of E2}
We start with an essentially pure Bose-Einstein condensate (BEC) of 1$\times$$10^5$ Cs atoms in a crossed optical dipole trap (XODT) adiabatically loaded into a cubic optical lattice with a lattice depth of $V_x=V_y=V_z=20$ $E_\mathrm{R}$, with $\lambda=1064.5\,\mathrm{nm}$ the wavelength of the lattice light. We take care to prepare a pure singly occupied Mott-insulator state by adapting the XODT during loading. During BEC preparation and lattice loading we apply a magnetic field gradient of 31 G/cm to compensate the gravitational force and we set the magnetic offset field $B$ to values around 21 G, giving a s-wave scattering length $a_s=210$ $a_0$. The field $B$ is then ramped to its final value with a resolution of $\sim8$ mG and a ramp speed of $~25\,\mathrm{G/ms}$. After a hold time of \un{50}{ms} we ramp $B$ to its final value of 18.5 G at maximum speed, switch off the lattice and XODT beams within 2 ms, and subsequently take an absorption picture of the atom cloud after an expansion time of \un{20}{ms} to determine the number of atoms. The lattice depth $V_q$ is calibrated via Kapitza-Dirac diffraction. The statistical error for $V_q$ is 1\%, though the systematic error can reach up to 5\%. The splitting of $P_1$ into $P^{'}_{1A}$ and $P^{'}_{1B}$ is \un{30}{a_0}. Assuming the theory from M2, such a splitting corresponds to an anisotropy of $\sim$1 $E_\mathrm{R}$, which is compatible with the systematic error for the lattice calibration. The scattering length $a_\mathrm{s}$ is calculated via its dependence on $B$ with an estimated uncertainty of \un{\pm5}{mG} arising from magnetic field noise and shot-to-shot variations. The resonances are further broadened by the magnetic field gradient. An estimate of the broadening can be obtained from the size of the cloud in the lattice. For simplicity, we assume that all the atoms are distributed on a sphere with perfect single-occupancy at the sites of the lattice. The levitation gradient of \un{31}{G/cm} gives then the variation of the field across the cloud, which can be converted to a variation $\Delta a_\mathrm{s}$ of $a_s$ according to Fig.~1(a) in the main text. For $P^{'}_\mathrm{1A}$, $P^{'}_\mathrm{1B}$, $P^{'}_\mathrm{2}$, and $P^{'}_\mathrm{3}$ we obtain 8, 13, 37, and $150\,a_0$, respectively. These variations are in fair agreement with the width of the resonances shown in Fig.~3(b) in the main text.

\subsubsection{Error bars for Fig.~4 in the main text}
The data in Fig.~4 in the main text are obtained by fitting data that is taken in the same way as the data shown in Fig.~3(a) in the main text. In view of limited sampling of the Gaussian fitting curves, the error bars are calculated from the sigma of the Gaussian fit. On top of that, we include the uncertainty from the lattice and magnetic-field calibration. The magnetic field is calibrated by microwave spectroscopy and the use of the Breit-Rabi formula, the uncertainty for E2 is $\Delta B=0.8\, G$. The lattice depth is calibrated via parametric heating measurements by driving the transition between the lowest and second excited band of the lattice near 20 $E_\mathrm{R}$, the uncertainty is $\Delta V=0.3\, E_\mathrm{R}$. Other systematic errors such as the ones deriving from drifts of the wavelength of the lattice laser and slight misalignment of the lattice beams are found to be negligible. 

\subsection{P4 discussion}
\subsubsection{Beyond the Bose-Hubbard model}
The standard Bose-Hubbard model is commonly derived under the assumption that the bandgap is the dominating energy scale of the system and that no mixing between the different bands takes place. When driving the superfluid-to-Mott-insulator transition, this leads to atoms confined in the $\psirelt\phicm(0,0,0)$ state. To assess the validity of the hypothesis the bandgap for different bands for non-interacting atoms in the lattice is compared to the on-site interaction energy calculated on the basis of the non-interacting Wannier functions. For a lattice with $V_0=20\,E_\mathrm{R}$, we find that difference in energy between the maximum of the lower band and the minimum of the first, second, and third excited band is $\Delta_1=7.6~E_\mathrm{R}$, $\Delta_2=13.2 E_\mathrm{R}$, and $\Delta_3=17.3~E_\mathrm{R}$, respectively, while for $a_\mathrm{h}=1500\,a_0$ the onsite interaction is $U=5.9~E_\mathrm{R}$. Thus, in the regime in which we are observing the resonances, the onsite interaction energy is non-negligible with respect to the bandgap. Noteworthy, a not fully adiabatic variation of the scattering length may cause an excitation to higher bands via avoided crossings~\cite{Schneider2009}.

\bibliography{ICIRs}

\end{document}